# Quantum polarization transformations in anisotropic dispersive medium

Yu I Bogdanov[1], A A Kalinkin[2,3], S P Kulik[4], E V Moreva[2,5], V A Shershulin[1,2,6]

1 Institute of Physics and Technology, Russian Academy of Sciences, 117218, Moscow, Russia
2 International Laser Center of Moscow State University, 119992, Moscow, Russia
3 Zavoisky Physical-Technical Institute, Russian Academy of Sciences, 420029, Kazan, Russia
4 Faculty of Physics, Moscow State University, 119992, Moscow, Russia
5 National Research Nuclear University "MEPHI", 115409, Moscow, Russia
6 National Research University of Electronic Technology MIET, 124498, Moscow, Russia

E-mail: **bogdanov_yurii@inbox.ru**, **sergei.kulik@gmail.com**

**Abstract.** Based on the concept of $\chi$-matrix and Choi-Jamiolkowski states we develop the approach of quantum process reconstruction. Special attention is paid to the adequacy of applied reconstruction models. The approach is applied to the statistical reconstruction of the polarization transformations in anisotropic and dispersive media realized by means of quartz plates and taking into account spectral shape of input states.

- PACS: 03.65.Wj, 03.67.Bg, 42.50.Dv

## 1. Introduction

One of the most important trends in the development of quantum information technologies associated with the development of a proper methodology for the control of quantum states and processes. This methodology is based on quantum measurements and is designed to provide an interface between the development of hardware components of quantum computers and its practical implementation.

From a mathematical point of view, the basis of this methodology is the quantification of the statistical theory of quantum operations and measurements, which is based on probabilistic Positive Operator Valued Measures (resolution of identity) and "completely positive mapping of operator algebras in Hilbert space" [1, 2]. From technological point of view such mathematical theory has to support methods, algorithms and software that must be able to provide adequate and complete assessment of the quality and effectiveness of specific quantum information systems [3, 4].

Quantum transformations formalism is used to describe reduced dynamics of open quantum systems. The base of such dynamics is a concept of complete positivity which was suggested and investigated in a number of works (K. Kraus [5], G. Lindblad [6], V. Gorini at al [7], D. Evans and J. Lewis [8] in statistical mechanics and A. Holevo [9] in quantum communication theory).

It is important to stress that the concept of complete positivity could be formalized through a variety of almost equivalent procedures, including: (i) based on the extended unitary dynamics of open quantum system interacting with an environment; (ii) through the Kraus operator product expansion, (iii) based on the Choi-Jamiolkowski isomorphism, as well as based on the formalism of quantum Markovian semigroups dynamics.

It is worthy to note that for the full and complete implementation of the problems of mathematical-modeling of quantum information technology is not enough to use any one of these approaches, and to use all these methods to describe their close relationship. Thus, the unitary representation of quantum operations in the extended space is needed to describe the relaxation processes in parallel with the evolution of the Hamiltonian. In this case, all processes are considered from the general theoretical positions on the basis of a single Schrödinger equation. The formalism of the operator-sum allows one to visualize the amount of decomposed non-unitary evolution of the density matrix of components, which are determined by the corresponding Kraus operators. Note, however, that the Kraus operators themselves are not uniquely defined (up to a very broad arbitrary unitary). Calculation of the Choi-Jamiolkowski states makes it obvious that the condition of complete positivity is satisfied. The relevant formalism is extremely important to analysis of quality of designed quantum gates. In fact, in this case, the analysis of evolution of an infinite number of possible states can be replaced by the study of a single state, though set in a space of higher dimension (if the channel acts in *s*-dimensional Hilbert space, then the corresponding Choi-Jamiolkowski state is specified in the $s^2$-dimensional space). Finally the evolution matrix is a very important notion: it simply and visually links input and output density matrixes. Note that $\chi$-matrix of Choi-Jamiolkowski and evolution matrix can be easily transformed one to another.

The paper is organized as following. It is basically divided in two parts. The first one relates to the theory (section 2-5) while the second one includes description and analysis of experiments (section 6-8). Sections 2 and 3 introduce general concept of $\chi$-matrix, Choi-Jamiolkowski states and based on these cornerstones quantum process tomography (QPT). Section 4 specifies application of the described above concepts to the polarization transformations performed by retardant plates on polarization states of light taking into account their dispersive properties. In section 5 we discuss results of numerical simulations for QPT performed with several popular protocols and pay special attention to the adequacy of applied reconstruction models. Section 6 presents experimental techniques for reconstruction of polarization transformations in anisotropic and dispersive objects (quartz plates) while section 7 describes the mixed state reconstruction as a sum over its quasi-pure components. Final original section 8 is devoted to the influence of instrumental uncertainties due to artificial optical anisotropy on the polarization reconstruction which seems to be important from practical problems of state preparation and measurement. Sections 9 and 10 conclude the paper.

## 2. Quantum transformations and quantum noise

It is the well-known fact that the ideal quantum gate is described by a unitary transformation of quantum state density matrix:

$$\rho_{out} = U \rho_{in} U^+ . \tag{1}$$

Real state evolution is never being unitary. In realistic models it is necessary to take into account unavoidable interaction of quantum system with environment (i.e. quantum noise). In the frame of theory of open quantum systems the state evolution is determined by operator sum $\mathbf{E}(\rho) = \sum_k E_k \rho E_k^\dagger$ [1,2], therefore the link between input and output states is given by

$$\rho_{out} = \sum_k E_k \rho_{in} E_k^+ , \tag{2}$$

where $E_k$ - are transformation elements (so called Kraus operators).

If the Hilbert space has dimension *s* then operators $E_k$ can be presented by $s \times s$-dimensional matrices. In the case of unitary transformation the operator sum has only single component, given by operator *U*.

Let transformation (2) conserve the trace of the density matrix, then the following equality holds

$$Tr(\rho_{out}) = Tr\left(\sum_k E_k \rho_{in} E_k^+\right) = Tr\left(\left(\sum_k E_k^+ E_k\right)\rho_{in}\right) = 1 . \tag{3}$$

Here we used invariance of the trace operation under cyclic permutation of its arguments. This equation holds for any input density matrices $\rho_{in}$. It is possible if transformation operators $E_k$ satisfy to the following normalisation condition:

$$\sum_k E_k^+ E_k = I , \tag{4}$$

where $I$ is an $s \times s$-dimensional identity matrix.

The $\chi$-matrix can be easily constructed using transformation elements $E_k$. This matrix plays a key role in quantum process tomography [10-12].

Let take the matrix $E_1$ with dimension $s \times s$ and stretch it into the column $e_1$ with length $s^2$ putting the second column under the first one and so on. This column will serve as a first column of some matrix $e$. Similarly, matrix $E_2$ gives the second column of matrix $e$ and so forth.

Let's define matrix $\chi$ based on the matrix $e$:

$$\chi = ee^+ . \tag{5}$$

By definition the matrix $\chi$ has dimension $s^2 \times s^2$ and it is important that $\chi$ can be interpreted as some density matrix in the $s^2$-dimensional space.

Then any quantum transformation reduces to some state in Hilbert space of higher dimension which is the so called Choi-Jamiolkowski isomorphism [1]. The corresponding state can be regarded as a joint state of two subsystems (each of dimension $s$). To account for the normalization condition (conservation of the trace) we must ensure that that the reduced matrix $\chi_A$ obtained by tracing out subsystem $B$, should be equal to the $s \times s$-dimensional identity matrix:

$$\chi_A = Tr_B(\chi) = I . \tag{6}$$

The $\chi$-matrix plays the same role as a density matrix in calculation of probabilities of possible measurement outcomes. Indeed let the input state be a pure state given by state-vector $|c_{in}\rangle$ and corresponding density matrix $\rho_{in} = |c_{in}\rangle\langle c_{in}|$. Note that at the same time the density matrix $\rho_{in}$ serves as a projector since $\rho_{in}^2 = \rho_{in}$. As a result of quantum transformation a new state appears at the output $\rho_{out}$. Let the measurement acts as a projection $\Pi = |c_m\rangle\langle c_m|$ of the output state onto the column-vector $|c_m\rangle$. Then as a consequence of Born-von Neumann rule the probability of measurement outcome is:

$$P = tr(\rho_{out}\Pi) . \tag{7}$$

As a next step let's consider the projection measurement onto equivalent effective state given by tensor product of complex conjugate of input state and the state responsible for the measurement outcome:

$$|\widetilde{c}_m\rangle = |c_{in}^*\rangle \otimes |c_m\rangle . \tag{8}$$

Corresponding projector is $\widetilde{\Pi} = |\widetilde{c}_m\rangle\langle\widetilde{c}_m|$. If one considers the $\chi$-matrix as some density matrix then the probability for equivalent effective measurement is:

$$\widetilde{P} = tr(\chi\widetilde{\Pi}) . \tag{9}$$

The straightforward calculation shows that both considered probabilities coincide $\widetilde{P} = P$. Therefore from probabilistic point of view quantum process is completely described by introduced $\chi$-matrix which is completely equivalent to assignment of transformation elements set $E_k$.

This property can be expressed in another equivalent form by using some auxiliary system (ancilla) and so called "relative-state" or Choi-Jamiolkowski state [1, 2, 12-15]. Let the quantum transformation $\mathbf{E}$ under consideration act on the $s$-dimensional system $A$. Then let's add the auxiliary system $B$ with the same dimension and consider the join system $AB$ which will submit to the input the maximally entangled state

$$|\Phi\rangle = \frac{1}{\sqrt{s}}\sum_{j=1}^{s}|j\rangle \otimes |j\rangle . \tag{10}$$

It is just the appropriate Choi-Jamiolkowski state. Here the first factor in the tensor product relates to the subsystem $B$ while the second one relates to the subsystem $A$. Let the identity transformation $\mathbf{I}$ be applied to the subsystem $B$. Then the transformation $(\mathbf{I} \otimes \mathbf{E})$ will be fulfilled in the join system $AB$. It turns out that the normalised $\chi$-matrix automatically appears at the output of the system if the density matrix of the input state is $|\Phi\rangle\langle\Phi|$:

$$(\mathbf{I} \otimes \mathbf{E})(|\Phi\rangle\langle\Phi|) = \rho_\chi, \tag{11}$$

where $\rho_\chi = \frac{1}{s}\chi$. The validity of the (11) can be checked by straightforward calculation:

$$\rho_\chi = \frac{1}{s}\sum_{j,j_1,k}|j\rangle\langle j_1| \otimes E_k|j\rangle\langle j_1|E_k^+ = \frac{1}{s}\chi \tag{12}$$

Figure 1 clarifies the argumentation presented above.

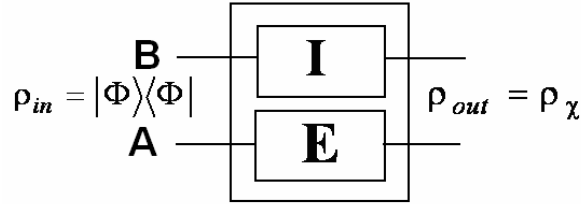

Figure 1. Quantum scheme for calculation of corresponding Choi-Jamiolkowski state.

In the following let us to designate as $\chi$ any $\chi$-matrix independently on its normalization. Up to now we have considered the construction of the $\chi$-matrix based on the transformation elements $E_k$. However it is easy to solve the inverse problem, namely finding $E_k$ knowing $\chi$. To do this let's diagonalize the $\chi$-matrix:

$$\chi = U_\chi D_\chi U_\chi^+. \tag{13}$$

Here $D_\chi$ is the diagonal matrix; its diagonal is formed by eigenvalues of $\chi$-matrix. All these eigenvalues are non-negative because of positive-definiteness of $\chi$-matrix. Let's settle eigen values in descending order. Columns of matrix $U_\chi$ are eigen vectors of $\chi$-matrix. Then the matrix $e$ can be found by formula:

$$e = U_\chi D_\chi^{1/2}. \tag{14}$$

Let the rank $r$ of quantum operation be the number of non-zero eigenvalues of $\chi$-matrix. Obviously $1 \leq r \leq s^2$. Thus the arbitrary quantum operation can be reduced to the form including not more than $s^2$ matrices $E_k$. The first rank case corresponds to unitary transformation.

Matrix $D_\chi$ can be easily reduced to the dimension $r \times r$ just by throwing away zero lines and columns. The matrix $U_\chi$ should be transformed to the dimension $s^2 \times r$ keeping first $r$ columns and throwing away the rest ones. Then the matrix $e$ takes the dimension $s^2 \times r$. In this case will continue to be carried out equality $\chi = U_\chi D_\chi U_\chi^+ = ee^+$.

Notice that matrix $e$ and correspondingly matrices $E_k$ are ambiguously determined. Let the matrix $e$ has $m$ columns and correspondingly its dimension is $s^2 \times m$ (with $m$ transformation elements $E_k$, $k = 1,2,...,m$). Obviously the $\chi$-matrix is not changed under the following transformation:

$$e \to e' = eU, \tag{15}$$

where $U$ is unitary $m \times m$-dimensional matrix.

New matrices $E'_k$ which are unitary equivalent to the set of initial matrices $E_k$ will correspond to new matrix $e'$. Notice that due to optimization procedure (which is simply reducing number of transformation elements) the minimum $m = r$ can be reached. Initial number of transformation elements $m$ can be even greater than $s^2$ (in principle it can be arbitrary large). It is important that the response of quantum system always can be described with not more than $s^2$ transformation operators independently on the number of elementary "noisy" operators $E_k$. This property reflects the important feature of finite-dimensional quantum systems, namely its limited informational nature. $\chi$-matrix can be assigned in different representations which are determined by different sets of basic matrices. In fact up to now we implicitly used the representation, which can be called "natural". Let's describe this representation evidently.

Let $|j\rangle$ be the ket-vector (column) with $j$-th element being equal to unit while the rest ones being zero. Analogously let $\langle k|$ be the bra-vector (row) with $k$-th element being equal to unit while the rest ones being zero. The matrix $|j\rangle\langle k|$ has unit element if it is settled on the cross of $j$-th row and $k$-th column. All the rest elements are zeros. If the indexes $j$ and $k$ take the values from 1 through $s^2$ ($j,k = 1,2,...,s^2$) then there are $s^4$ of such matrices. It is obvious that the $\chi$-matrix can be decomposed in the following form:

$$\chi = \sum_{j,k} \chi_{jk} |j\rangle\langle k| . \qquad (16)$$

Here the set of $s^4$ matrices $|j\rangle\langle k|$ plays the role of the basis. Of course the decomposition coefficients change under transformation from matrices $|j\rangle\langle k|$ to other basis sets of matrices. It corresponds to another representation of the $\chi$-matrix.

As an example let's describe the transformation from the "natural representation" introduced above to another widely used representation, given by Pauli matrices.
The single-qubit basis set, determined by four $2 \times 2$ matrices constructed by unit matrix $E$ and Pauli matrices $X, Y, Z$ has the form:

$$I = E/\sqrt{2}, X = \sigma_x/\sqrt{2}, Y = -i\sigma_y/\sqrt{2}, Z = \sigma_z/\sqrt{2} . \qquad (17)$$

Then the two-qubit basis set is formed by tensor products of all possible pairs of those matrices (totally 16 operators), the three-qubit basis set is formed by 64 operators and so forth.

The transformation from matrix $\chi$ in "natural" basis to matrix $\chi'$ in the basis given by Pauli matrices is determined by following unitary transformation

$$\chi' = U_0^+ \chi U_0 . \qquad (18)$$

Basic matrices (17) or their tensor products stretched into the columns determine the columns of corresponding unitary transformation matrix $U_0$.

It is worth mentioning that considered sets of basis matrices are orthonormal. In general, the set of basis matrix ($j = 1,...,m$) is an orthonormal if

$$Tr(a_j a_k^+) = \delta_{jk} \qquad j,k = 1,...,m . \qquad (19)$$

## 3. Quantum process tomography

Quantum process tomography is equivalent to statistical reconstruction of Choi-Jamiolkowski state $\rho_\chi$ (11). Performing quantum measurement protocol consisting on $m$ rows [16] one registers $m$ values of frequency of events $k_j$, $j = 1,...m$. If the exposure time of $j$-th row of the protocol equals $t_j$, then the registered in experiment event number $k_j$ is an random variable characterized by Poisson distribution with mean value $\lambda_j t_j$:

$$P(k_j) = \frac{(\lambda_j t_j)^{k_j}}{k_j!} \exp(-\lambda_j t_j) . \qquad (20)$$

Here the intensity of the event generation $\lambda_j$ (or the expected number of events while using the registration scheme measuring the event rate) is determined by corresponding intensity operator $\Lambda_j$:

$$\lambda_j = Tr(\Lambda_j \rho_\chi). \qquad (21)$$

The most convenient parameterisation for the $\chi$-matrix and corresponding density matrix $\rho_\chi$ is acquired by means of purification procedure, which is determined by (5). Matrix $e$ after proper normalization corresponds to the purified state-vector $c$. Notice that due to the unitary arbitrariness (15) the state vector $c$ is ambiguously determined. However all possible state-vectors correspond to the same density matrix $\rho_\chi$. Using purified state-vector the formula (21) takes the form:

$$\lambda_j = \langle c | \Lambda_j | c \rangle. \qquad (22)$$

Each row of the protocol corresponds the concrete set of the measurement parameters, which corresponds to selection of some projection (8) of the quantum state at the input and output. In this paper we used three different families of protocols introduced earlier in the works ("R"(Singapore)-family: [17, 18]; "J"-family: [19, 20]; "BN"(Moscow)-family: [3, 4, 16]). The protocol J4 suggested in [19] performs projective measurements of (polarization) qubits upon fixed components of the Stokes vector: $|H\rangle$, $|V\rangle$, $|-45^0\rangle = \frac{1}{\sqrt{2}}\{|H\rangle - |V\rangle\}$, and $|L\rangle = \frac{1}{\sqrt{2}}\{|H\rangle - i|V\rangle\}$. In experiment these measurements usually are performed using two fixed retardant plates (half-wave and quarter-wave ones) and polarization prism selecting particular (vertical) polarization. If the measured qubit were projected on the states possessing tetrahedral symmetry then the protocol transforms to R4. The over-complete BN protocol exploits a single retardant plate and fixed polarization prism. The corresponding measurements are performed for $N$ orientations of the plate with a step of $180^0/N$. Optimization of such measurements is described in details in [3, 4].

The goal of statistical reconstruction of quantum process is finding the best (in some sense) reconstruction of the corresponding Choi-Jamiolkowski state by proper processing of experimental data. One of the best methods providing such reconstruction is Fisher maximal likelihood method. The task of the present paper is finding such purified state-vector $c$, which gives the maximum of likelihood function. In our case this function corresponds to the product of Poisson probabilities over all rows of the protocol:

$$L = \prod_{j=1}^{m} \frac{(\lambda_j t_j)^{k_j}}{k_j!} \exp(-\lambda_j t_j). \qquad (23)$$

Necessary condition for an extremum of the function (23) leads to likelihood equation [16]:

$$Ic = Jc, \qquad (24)$$

where $I$ and $J$ - so called theoretical and empirical Hermitian Fisher information matrices:

$$I = \sum_{j=1}^{m} t_j \Lambda_j, \quad J = \sum_{j=1}^{m} \frac{k_j}{\lambda_j} \Lambda_j. \qquad (25)$$

Normalization condition, which is automatically contained in likelihood equation (24) has the form:

$$\sum_{j=1}^{m} \lambda_j t_j = \sum_{j=1}^{m} k_j = n, \qquad (26)$$

where $n$ is a total number of registered events.

Condition (26) links the total number of registered events $n$ with the total (over all rows of the protocol) expected number of events. In the approach developed in the present paper condition (26) replaces usually used identity normalization condition $\langle c | c \rangle = 1$. However the state reconstructed by quantum tomography state would be the true Choi-Jamiolkowski state if matrix normalization condition (4), (6) were fulfilled. For the resulting quantum state to be the true state of Choi-Jamiolkowski, it must still satisfy the matrix normalization conditions (4), (6).

Corresponding condition leads to the fact that reduced (with respect to input A) Choi-Jamiolkowski state has to be completely mixed and be described by density matrix:

$$\rho^{(A)} = \frac{I_s}{s}, \qquad (27)$$

Where $I_s$ is the $s$-dimensional identity matrix. Considered normalization condition can be taken into account by means of additional constraints imposed on likelihood function (23) [15]. However in the present paper we chose alternative way consisting in adding auxiliary statistics. This trivial auxiliary statistics corresponds to virtually measured states (27). Notice that for completely mixed state $\rho^{(A)}$ being projected onto any state vector $|c_{in}^*\rangle$ one gets the same outcome probability $\frac{1}{s}$:

$$P(c_{in}^*) = Tr(|c_{in}^*\rangle\langle c_{in}^*|\rho^{(A)}) = \frac{1}{s}. \qquad (28)$$

Corresponding measurement operator in the Choi-Jamiolkowski $s^2$-dimensional space is a summation over all possible outcomes at the output (in the subsystem $B$):

$$\Lambda(c_{in}^*) = |c_{in}^*\rangle\langle c_{in}^*| \otimes I_s. \qquad (29)$$

Here the Choi-Jamiolkowski state $\rho_\chi$ plays a role of a measurable state and

$$Tr(\rho_\chi \Lambda(c_{in}^*)) = \frac{1}{s}. \qquad (30)$$

Relation (30) determines one of auxiliary rows for each $|c_{in}^*\rangle$. We also assume that the set of states $|c_{in}^*\rangle$ is tomographically complete [21].

Accuracy of maximal likelihood method is determined by matrix of complete information, which is an analogue of Fisher information matrix

$$H = 2\sum_j \frac{t_j(\Lambda_j c)(\Lambda_j c)^+}{\lambda_j}, \qquad (31)$$

used for estimation of the quantum state [22].

Matrix (31) is assigned in real Euclidean space of doubled dimension. To extract state-vector $c$ in this representation one needs to settle the imaginary part of purified state vector right under its real part.

**4. Calculation of $\chi$-matrix for retardant dispersive plate**

Ideal unitary transformation performed by retardant plate on monochromatic light is rotation on the Poincaré - Bloch sphere:

$$U = exp(-i\delta\vec{\sigma}\vec{n}) = I\cos\delta - i\vec{\sigma}\vec{n}\sin\delta =$$
$$\begin{pmatrix} \cos\delta - in_z\sin\delta & -i(n_x - in_y)\sin\delta \\ -i(n_x + in_y)\sin\delta & \cos\delta + in_z\sin\delta \end{pmatrix}. \qquad (32)$$

This rotation can be written as follows:

$$U = exp(-i\delta\{n_x\sigma_x + n_z\sigma_z\}), \qquad (33)$$

where $\delta = \frac{\pi \Delta n h}{\lambda}$ is an optical thickness of the plate, $\Delta n = n_o - n_e$ is the birefringence of the plate material at a given wavelength $\lambda$, and $h$ is its geometric thickness, $n_x = sin(2\alpha)$, $n_z = cos(2\alpha)$, $\alpha$ is the orientation angle between the optical axis of the phase plate and vertical direction $z$.

We assume that light propagates along axis Y while retardant plate and its optical axis lie in the plane XZ.

For quartz plates which we used in experiment the dispersion of the refractive indexes is given by dispersion equations [23]:

$$n_0 = \sqrt{1.30979 + \frac{1.04683 \cdot \lambda^2}{\lambda^2 - 0.01025} + \frac{1.20328 \cdot \lambda^2}{\lambda^2 - 108.584}} \qquad (34)$$

$$n_e = \sqrt{1.32888 + \frac{1.05487 \cdot \lambda^2}{\lambda^2 - 0.01053} + \frac{0.97121 \cdot \lambda^2}{\lambda^2 - 84.261}}, \qquad (35)$$

where the wavelength $\lambda$ is supposed to be given in micrometers.

Let's convert unitary transformation (32) to state-vector form. To do this we stretch the $2 \times 2$ matrix into the column with length 4 and normalize it:

$$|\Psi\rangle = \frac{1}{\sqrt{2}} \begin{pmatrix} \cos\delta - in_z \sin\delta \\ -i(n_x + in_y)\sin\delta \\ -i(n_x - in_y)\sin\delta \\ \cos\delta + in_z \sin\delta \end{pmatrix}. \qquad (36)$$

This vector is defined by single parameter $\delta$. Due to dispersion different wavelengths $\lambda_j$ correspond to different optical thickness $\delta_j$ and rotation angles $\theta_j = 2\delta_j$ on Poincaré - Bloch sphere, i.e different state-vectors $|\Psi_j\rangle$ while the rotation axis $\vec{n} = \{n_x, n_y, n_z\}$ does not change.

It is worth to point out that these states form incoherent mixture with weights $P_j = P(\lambda_j)\Delta\lambda$, where $P(\lambda)$ is a spectral density and $\Delta\lambda$ is a wavelength quantization step. In order to reach as high accuracy as possible one needs to substitute the continuous wavelength interval with large enough number of steps (knots). In the present work we chose about 800 knots for the whole wavelength interval under numerical calculations and 7 knots under experimental procedure (see below).

It is easy to notice that in fact the vector $|\Psi\rangle$ is assigned in two-dimensional subspace

$$|\psi\rangle = a|\varphi_1\rangle + b|\varphi_2\rangle \qquad (37)$$

with orthonormal basis states:

$$|\varphi_1\rangle = \frac{1}{\sqrt{2}}\begin{pmatrix} 1 \\ 0 \\ 0 \\ 1 \end{pmatrix}, \quad |\varphi_2\rangle = \frac{1}{\sqrt{2}}\begin{pmatrix} n_z \\ n_x + in_y \\ n_x - in_y \\ -n_z \end{pmatrix}. \qquad (38)$$

Hence the probability amplitudes take the following form:

$$a = \cos(\delta) \quad b = -i\sin(\delta). \qquad (39)$$

Then the density matrix for the given wavelength is:

$$\rho = |\psi\rangle\langle\psi| = aa^*|\varphi_1\rangle\langle\varphi_1| + bb^*|\varphi_2\rangle\langle\varphi_2| + ab^*|\varphi_1\rangle\langle\varphi_2| + ba^*|\varphi_2\rangle\langle\varphi_1|. \qquad (40)$$

Therefore the density matrix of the mixed states, corresponding to $\chi$-matrix is:

$$\rho_\chi = \sum_j |\psi_j\rangle\langle\psi_j| P_j. \qquad (41)$$

Finally we get the $\chi$-matrix:

$$\chi = 2\rho_\chi. \qquad (42)$$

The consideration above clearly show why the $\chi$-matrix of a single waveplate has a rank $r = 2$ while in general case it can be $r = 4$.

## 5. Numerical experiments

For numerical calculations we used the following parameters: central wavelength is $\lambda = 1.1509\,\mu m$, the spectrum shape is given by function $\left[\dfrac{sin(x)}{x}\right]^2$, Full Width Half Maximum (FWHM) is $\Delta\lambda = 0.008\,\mu m$, geometrical thickness of the retardant plate is $h = 5024\,\mu m$, orientation angle is $\alpha = \pi/4$.

In this case the density matrix for the Choi-Jamiolkowski state to be reconstructed is:

$$\rho_\chi = \begin{pmatrix} 0.42099 & -0.0047933\,i & -0.0047933\,i & 0.42099 \\ 0.0047933\,i & 0.079005 & 0.079005 & 0.0047933\,i \\ 0.0047933\,i & 0.079005 & 0.079005 & 0.0047933\,i \\ 0.42099 & -0.0047933\,i & -0.0047933\,i & 0.42099 \end{pmatrix}. \quad (43)$$

This density matrix has the rank $r = 2$ and its non-zero eigenvalues are $\lambda_1 = 0.84212$, $\lambda_2 = 0.15788$. The final purpose of the following research is to compare the accuracy of the quantum process reconstruction for two selected protocols J4 and R4. We are also going to demonstrate the non-adequacy of the suggested procedure which does not take into account the incomplete rank of the quantum process.

Numerical results for R4 protocol are presented on figures 2 and 3; results for J4 protocol are on figures 4 and 5. Figure 2 corresponds to quantum process reconstruction by means of adequate method when data generation and its reconstruction were done along the same rank $r = 2$ model. In this case one can see good agreement between numerical calculations and exact theoretical prediction of fidelity distribution. For R4 protocol we chose $n = 10^4$ in each experiment, while number of experiments is $N = 50$.

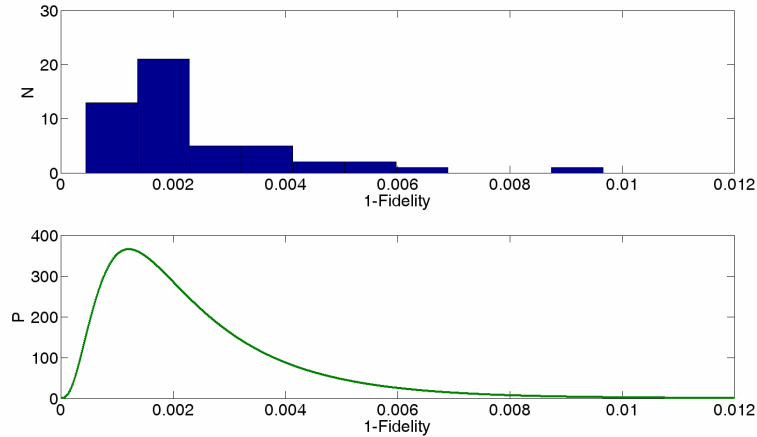

Figure 2. Protocol R4, adequate model. Numerical calculations (upper bar graph) and theoretical prediction for the fidelity distribution. Rank $r = 2$.

Figure 3 demonstrates the non-adequate reconstruction algorithm, when data generation is performed by means of model with rank $r = 2$, while reconstruction procedure has been done by model of rank $r = 4$. In this case it turns out that the accuracy losses are much higher than in corresponding adequate reconstruction method. As in previous case the reconstruction is performed by R4 protocol with $n = 10^4$ and $N = 50$. Detailed numerical calculations show that non-adequacy of the reconstruction model leads to increase of losses more than 6 times.

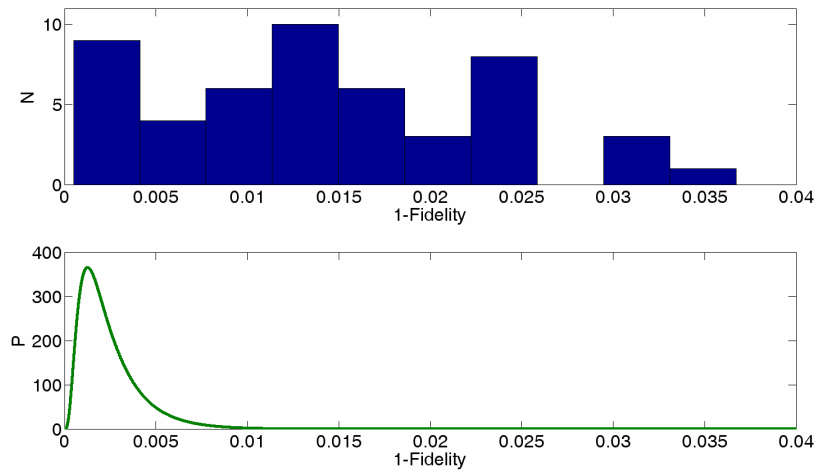

Figure 3. Protocol R4, non-adequate model. Numerical calculations (upper bar graph) and theoretical prediction (lower density) for the fidelity distribution. Model rank for the data generation is $r = 2$, model rank for the reconstruction is $r = 4$.

Figure 4 is analogous to figure 2 and corresponds to the adequate method of reconstruction with model rank $r = 2$ by means of J4 protocol. In this case there is a good agreement between numerical experiments and theoretical predictions. We used the same parameters of the protocol as for R4 $n = 10^4$ and $N = 50$. Again one can see that J4 protocol yields to R4 in the respect of accuracy (on average more than 1.6 times).

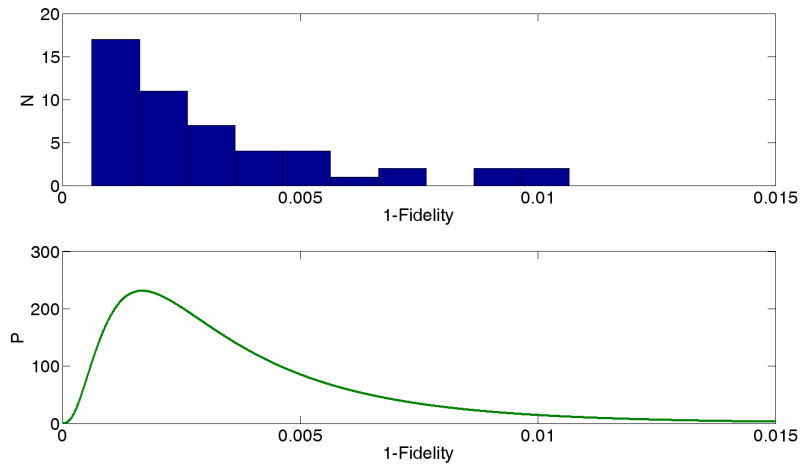

Figure 4. Protocol J4, adequate model. Numerical calculations (upper bar graph) and theoretical prediction for the fidelity distribution. Rank $r = 2$.

Finally the figure 5 demonstrates non-adequate reconstruction algorithm by means of J4 protocol, when data generation is performed by means of model rank $r = 2$, while reconstruction procedure has been done by means of model rank $r = 4$ with $n = 10^4$ and $N = 50$. As in the case of R4 protocol the accuracy losses is much higher than in corresponding adequate reconstruction method. Non-adequacy of the reconstruction model leads to increase of losses more than 8 times.

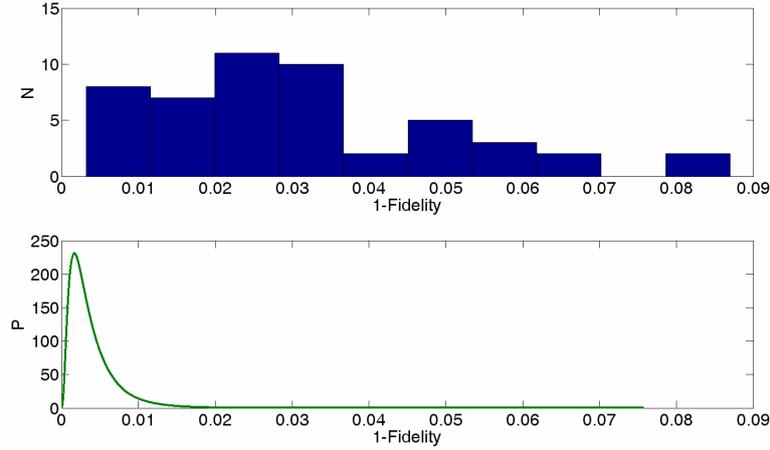

Figure 5. Protocol J4, non-adequate model. Numerical calculations (upper bar graph) and theoretical prediction for the fidelity distribution. Model rank for the data generation is $r = 2$, model rank for the reconstruction is $r = 4$.

The number of real parameters that define the $N$-qubit operation of rank $r$ in the Hilbert space of dimension $s = 2^N$ is $\nu = 2s^2 r - r^2 - s^2$. For full rank operation we have $r = s^2$, $\nu = s^4 - s^2$. For an operation of incomplete rank the corresponding number of degrees of freedom can be much smaller if $r \ll s^2$. If an incomplete rank quantum operation is reconstructed by a full rank model then that model is clearly inadequate. The study above shows that using such kind of an inappropriate model leads to a sharp decrease in the accuracy of statistical reconstruction of quantum operations. In fact, it turns out that asymptotically, as sample size $n$ increases, the loss of fidelity is proportional to $1/\sqrt{n}$ for the inadequate model (compared to $1/n$ for an adequate model) [24].

### 6. Experimental set-up and protocol of quantum measurements

The experimental set-up is shown on figure 6a. The halogen lamp serves as a source of wide-band light. Monochromator MDR-41 selects the radiation with central frequency 1.1509 $\mu m$ and variable spectral range from 0.8 up to 8 nm. Paraxial light beam is formed by single-mode fiber SMF-28 supplied with micro objectives F240FC-1550. Simultaneously such an optical system provides fixed attenuation of the photon flux, so the registration system works in proper linear regime. A Glan-Thompson prism is used for selecting vertical polarization $|V\rangle$, serving as initial state for the following transformations performed by two retardant plates with geometrical thicknesses $h_1 = 19.5 \mu m$ and $h_1 = 312.7 \mu m$ staying in series.

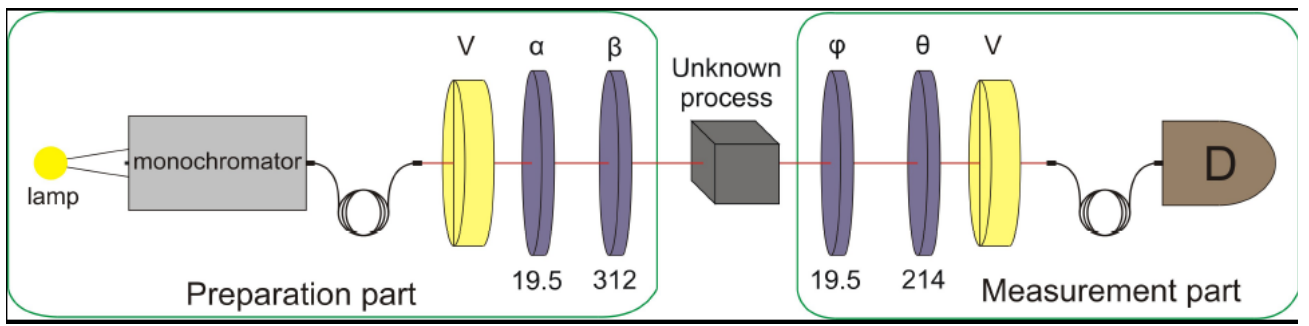

Figure 6a. The experimental set-up for quantum process tomography by protocols R4 and J4. Thick quartz plates simulated unknown quantum process by introducing polarization transformation at given frequency component of the input state(s).

These plates provide the set-up with three sets of given initial states. The first set "J4" corresponds to polarization states being components of the Stokes vector: $|H\rangle$, $|V\rangle$, $|-45^0\rangle = \frac{1}{\sqrt{2}}\{|H\rangle - |V\rangle\}$, and $|L\rangle = \frac{1}{\sqrt{2}}\{|H\rangle - i|V\rangle\}$ [19, 20]. The second set "R4" the polarization states are chosen in the way to satisfy the following conditions:

$$\vec{a}_j \cdot \vec{a}_k = \frac{4}{3}\delta_{jk} - \frac{1}{3} = \begin{cases} 1 & \text{if } j = k \\ -\frac{1}{3} & \text{if } j \neq k \end{cases}. \tag{44}$$

Corresponding four vectors are placed symmetrically on the Poincaré - Bloch sphere and form a tetrahedron [17].

In the third set "B4" the states under reconstruction are formed starting from initial $|V\rangle$ polarization by following rotation of retardant plate $h_1 = 214\mu m$ at angles $\alpha = 0^0$, $15^0$, $30^0$, $45^0$ (Figure 6b).

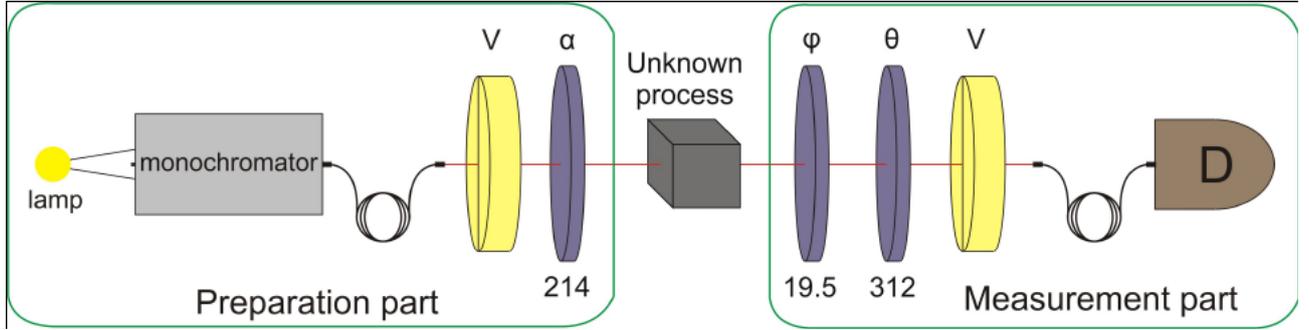

Figure 6b. The experimental set-up for quantum process tomography by protocol B4. Thick quartz plates simulated unknown quantum process by introducing polarization transformation at given frequency component of the input state(s).

The measurement part of the set-up consists of two retardant plates $h_3 = 19.5\mu m$ and $h_4 = 214\mu m$, followed by analyser (Glan-Thompson prism transmitting vertical polarization $|V\rangle$) and single-photon detector based on GaAs-based avalanche photo-diode with a pigtailed input and an internal gate shaper [25]. Each of three sets of polarization states has been measured in eigen basis. For instance, each of states from "J4" basis was reconstructed by means of tomographic protocol, based on projective measurements onto the corresponding states $|H\rangle$, $|V\rangle$, $|-45^0\rangle = \frac{1}{\sqrt{2}}\{|H\rangle - |V\rangle\}$, and $|L\rangle = \frac{1}{\sqrt{2}}\{|H\rangle - i|V\rangle\}$. Hence for each set we have measured 16 projections. Knowing the input and measured (output) states one can reconstruct $\chi$-matrix of the quantum process. In our experiments the role of quantum channel transforming the input states plays the set of thick quartz plates with thickness $5031\mu m$ each, whose optical axis is oriented at $45^0$ with respect to vertical direction. For each set of states "J4", "R4", and "B4" two series of experiments have been performed. In the first series the quantum channel was loaded with quasi-monochromatic radiation (bandwidth is 0.8 nm). For this sort of states different spectral components forming its spectrum get almost the same polarization transformations because they are practically coincide in frequency, so the state can be considered as a quasi-pure. In the second series the slit of monochromator was completely open, so the broadband radiation passed through (FWHM is about 8 nm). Passing through the thick quartz plates different spectral components get different polarization transformation due to dispersion, so the output state was quasi-mixed in polarization. Using single thick quartz plate provides with 63% of mixture in the output states while using two plates gives 98% of the mixture.

Comparison of $\chi$-matrices for quantum processes, reconstructed for each series of experiments with the $\chi$-matrices obtained by simulation of corresponding quantum processes are shown in table 1.

Table 1. Reconstruction of $\chi$-matrices of quantum processes.

| Number of plates | J4 protocol | | R4 protocol | | B4 protocol | |
|---|---|---|---|---|---|---|
| | Fidelity | | | | | |
| | Spectrum width (nm) | | | | | |
| | 0.8 | 8 | 0.8 | 8 | 0.8 | 8 |
| 0 | 0.9971 | 0.9938 | 0.9978 | 0.9971 | 0.9724 | - |
| 1 | 0.9541(0.9279) | 0.9812 | 0.9735 | 0.9749(0.8259) | 0.9679 | 0.9707 |
| 2 | 0.9486 | 0.9982 | 0.9669 | 0.9706 | 0.9426 | 0.9678 |

Generally the spectral structure of the input polarization state is not considered in quantum process tomography, so the whole state is supposed to be transformed as a pure one. However this is not true and sometimes the spectral structure of the state dramatically affects the procedure of reconstruction [24]. For example practically used biphoton states possess originally wide spectrum and without spectral post-selection it is not quite correct to consider such states as polarization pure states. Thus analysis of the polarization reconstruction process taking into account the spectral structure of the states is of interest both for quantum state and quantum process tomography.

Convenient comparison of statistical reconstruction performed by different protocols can be performed based on the universal statistical distribution of quantum tomography accuracy [3]. The table 1 presents the fidelities measured with protocols J4, R4, B4. Three different quantum processes for various spectral ranges 0.8 nm(quasi-pure) and 8 nm (mixed) and each of the protocol were reconstructed. The first row corresponds to identity transformation, when there are no any quartz plates between preparation and measurement parts. In this case the $\chi$-matrix of the quantum process is equal to unitary matrix with accuracy, depending on quality of state prepartion/measurement procedure itself. The second and third rows correspond to the reconstruction of the quantum processes occurring in the single or double thick quartz plates. The procedure of reconstruction depends strongly on the model rank. For the narrow enough spectrum of the input polarization states the quality of reconstruction is higher when the simple model rank is used. For example for the protocol J4 the quantum process in the single thick quartz plate with quasi-pure states is reconstructed with fidelity F=0.9541. If we reconstruct the same quantum process by the algorithm with higher rank, the fidelity becomes lower F=0.9279. The opposite situation is for the protocol R4, when the quantum process in the single thick quartz plate with mixed states takes place. Here the adequate model, which takes into account the spectral width gives the higher value of the fidelity F=0.9749 (and F=0.8259 for non-adequate model).

**7. Mixed state reconstruction as a sum over its quasi-pure components**

Corresponding experiments have been performed in three stages. At the first stage we measured mixed states with different degrees of mixture.

Arbitrary mixed state is given by the density matrix:

$$\rho = p_1 |H\rangle\langle H| + p_2 |V\rangle\langle V|, \qquad (45)$$

where the basic vectors can be formed as an action of creation operators in vertical and horizontal polarization modes on the vacuum state

$$|H\rangle = a^\dagger |vac\rangle \equiv \begin{pmatrix} 1 \\ 0 \end{pmatrix}, \quad |V\rangle = b^\dagger |vac\rangle \equiv \begin{pmatrix} 0 \\ 1 \end{pmatrix}. \qquad (46)$$

At the same time the corresponding commutator $[a,b]=0$, and real coefficients satisfy to normalization condition $\sum_{i=1}^{2} p_i = 1$. The state (45) is often interpreted as an averaging over classical probability distribution $p = (p_1, 1-p_1)$ of the vector state

$$|\Psi\rangle = c_1|H\rangle + c_2|V\rangle \qquad (47)$$

with random amplitudes $c_1$ and $c_2$ ($|c_1|^2 = p_1$, $|c_2|^2 = p_2$).

In experiment the mixed (or multimode) polarization light state with given spectral width $\Delta\omega = \dfrac{2\pi}{\tau_{coh}}$ can be formed by introducing the delay exceeding the coherence length $l > l_{coh} = \dfrac{2\pi c}{\tau_{coh}}$ between "H" and "V" components of initially pure polarization state (47).

In our experiments as an element introducing the delay we used birefringent material (quartz plate) oriented at $45^0$ with respect to basic polarization vectors $|H\rangle$ and $|V\rangle$ of the initial state (47).

The monochromator selects the central wavelength 1000 nm with spectral width 8 nm (see figure 6c).

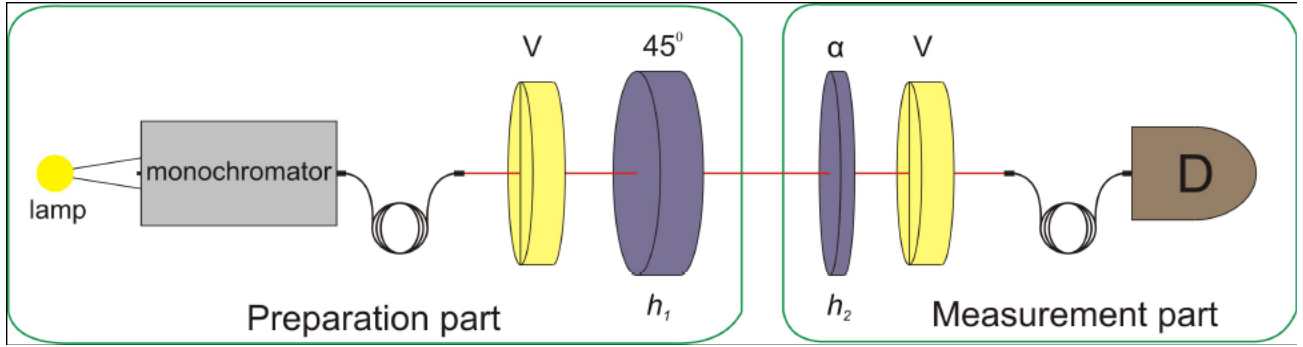

Figure 6c. Experimental set-up for component-wise reconstruction of mixed polarization states.

At the first stage we used the Glan-Tompson prism, selecting vertical polarization component, and set of thick quartz plates ($5031\mu m$) oriented at $45^0$ for preparation of the initial polarization state to be reconstructed. As a result initially pure broadband state is transformed to the mixed states with mixture degree depending on the spectral width of the prepared state. Indeed since the measuring part of the set-up posses finite spectral band the polarization components were integrated over this band and mixed polarization state was finally registered. In experiment two different states of polarization qubit were prepared with different degrees of mixture. For the statistical reconstruction of the state we have used B36 protocol to measure the event rate at each of 36 orientation of the single retardant plate $h_2$ with thickness $312.7\mu m$ introduced in the measurement part of the set-up. These measurements have been performed for several spectral components with varying transmission wavelength of the monchromator. In experiments we have prepared two polarization qubits with different degrees of mixture using single or doubled thick quartz plate(s) (see above). Then we applied maximal likelihood method developed in [24] to the obtained data for the statistical reconstruction of input states with different degrees of mixture.

Fidelity serves as a good quantitative measure for the correspondence between theoretical and reconstructed states. Generally fidelity takes a form:

$$F = \left(Tr\sqrt{\rho_0^{1/2} \rho \rho_0^{1/2}}\right)^2 \quad (48)$$

where $\rho_0$ and $\rho$ are theoretical and reconstructed density matrices correspondingly.

For numerical calculations we divide the initial spectrum of the input state into relatively small segments, while the polarization state is the sum of states corresponding to different segments:

$$|\Psi\rangle = \sum_k a_k |\Phi(\omega_k)\rangle \quad (49)$$

where $|\Phi(\omega_k)\rangle = c_1(\omega_k)|H\rangle + c_2(\omega_k)|V\rangle$ and amplitudes $a_k$ are determined by the spectral shape of the light passing through the monochromator.

Before passing transforming polarization elements (retardant plates) the density matrix takes the form:

$$\rho^{in} = |\Psi\rangle\langle\Psi| = \sum_{k,j} a_k a_j^* |\Phi(\omega_k)\rangle\langle\Phi(\omega_j)| \qquad (50)$$

As it was noted before the thin retardant plates transform the state as a whole, i.e. all spectral components get the same polarization transformation, whereas the thick retardant plate transforms the initial density matrix as

$$\rho' = G(\omega_k)\rho^{in}G^+(\omega_k),$$

where the matrix

$$G(\omega_k) = \begin{pmatrix} t_k & r_k \\ -r_k^* & t_k^* \end{pmatrix}, \qquad (51)$$

depends on complex coefficients $r_k = i\sin(\delta_k)\sin(2\alpha)$, $t_k = \cos(\delta_k) + i\sin(\delta_k)\cos(2\alpha)$. Here index $k$ denotes chosen frequency component.

Using (51) let's link each input and output frequency component and find the reduced density matrix taking the trace over frequencies:

$$\rho = \sum_k |a_k|^2 G(\omega_k)|\Phi(\omega_k)\rangle\langle\Phi(\omega_k)|G^\dagger(\omega_k). \qquad (52)$$

As a result we got fidelity for the first mixed state (63% mixture) $F_1 = 0.997$ and $F_2 = 0.9986$ for the second one (98% mixture).

At the second stage each of two mixed states was measured component wise. To do this we divided the whole spectrum having the shape of $\left[\dfrac{sin(x)}{x}\right]^2$ into seven ranges with central wavelengths $\lambda_i = 994nm,\ 996nm,\ 998nm,\ 1000nm,\ 1002nm,\ 1004nm,\ 1006nm$ as shown on the figure 7.

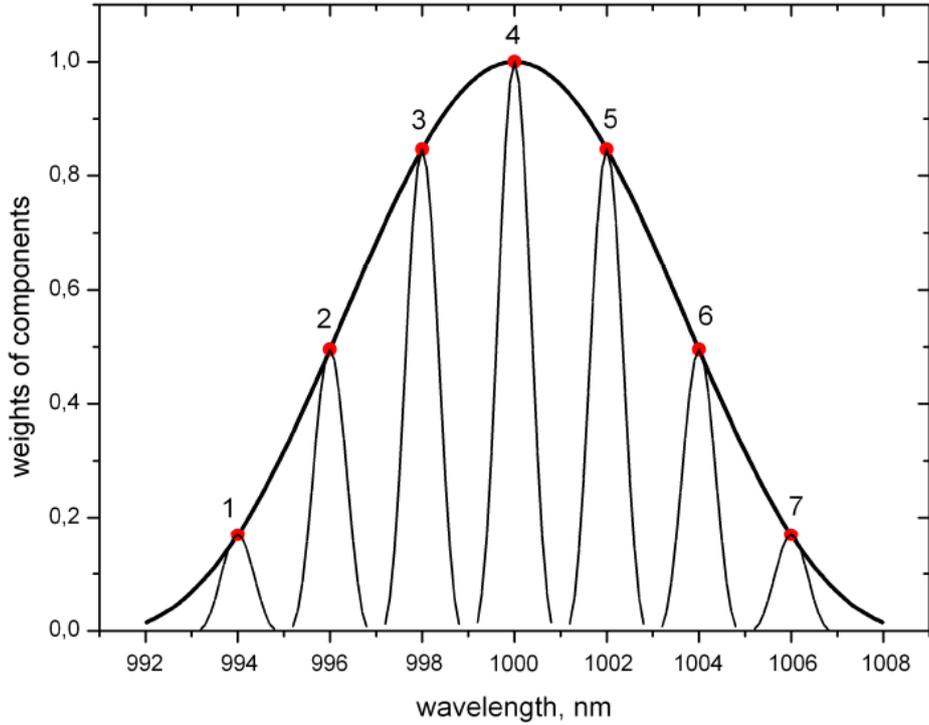

Figure 7. The solid line shows the spectral shape of the radiance passing through the monochromator (open slit). Numbered points designate the selected spectral components used for the spectrum digitization and corresponding spectral shape for the narrow slit of monochromator.

The experimental set-up was the same. However at this stage the monochromator selected relatively narrowband radiation 0.8 nm with indicated above central wavelengths in series. Polarization states at each of 7 spectral components have been reconstructed by means of B36 protocol. Corresponding wavelengths and calculated fidelities of (pure state) density matrix reconstruction are placed in the table 2.

**Table 2.** Reconstructed quasi-pure components for two mixed polarization states.

| Number of spectral component, $i$ | Fidelity, $F\left(\rho_{theor}^{pure}, \rho_{exp}^{i}\right)$ | | Central wavelength (nm) | Weight of the component, $a_i$ |
|---|---|---|---|---|
| | Single plate (63% mixture) | Double plate (98% mixture) | | |
| 1 | 0.9970 | 0.9983 | 994 | 0.1690 |
| 2 | 0.9995 | 0.9975 | 996 | 0.4955 |
| 3 | 0.9994 | 0.9947 | 998 | 0.8470 |
| 4 | 0.9984 | 0.9972 | 1000 | 1 |
| 5 | 0.9965 | 0.9940 | 1002 | 0.8470 |
| 6 | 0.9958 | 0.9961 | 1004 | 0.4955 |
| 7 | 0.9966 | 0.9972 | 1006 | 0.1690 |

In order to reconstruct the mixed states as a sum of its quasi-pure components we have used the following formula:

$$\rho_{\exp}^{comp} = \sum_{i=1}^{7} a_i \rho_i ,  \quad (53)$$

where $\rho_{\exp}^{comp}$ is the density matrix of the mixed state; $a_k$ are the weights of its components (see table 1), which are calculated using transmission function of the monochromator, shown on figure 7; $\rho_i$ are the density matrices of $i$-th quasi-pure states.

The table 3 contains the results of statistical reconstruction with different contributions of different spectral components. The first line shows the total fidelity if all spectral components contributed correspondingly.

At the third stage of experiments we compare the accuracy of reconstruction using formula (53) depending on number of selected quasi-pure components of the state/spectrum. The second and lower lines show value of fidelity if partial components are used for the reconstruction procedure. Notice that the fidelity of the reconstruction takes relatively high values ($F > 0.99$) even if small number of frequency components are involved. We have estimated the entropy of the resulting state as well as fidelities between theoretical density matrix $\rho_{teor}^{mix}$ and experimental ones calculated using (53) for different sets of quasi-pure components (see column 1) and both polarization qubits with different degrees of mixture. The last value can be extracted using simple formula for the entropy

$$S = -\sum_{n=1}^{2} \lambda_n \log \lambda_n, \tag{54}$$

where $\lambda_n$ are the eigenvalues of the density matrix.

The table 3 clearly demonstrates that even three components (for instance components 2, 4, 6) chosen symmetrically and uniformly with respect to the central wavelength of the spectral profile are sufficient for mixed state reconstruction with fidelity $F > 0.99$. For irregular choice of the components the quality of the reconstruction falls down which leads to increasing purity of the reconstructed state, which is more or less predictable result following from qualitative considerations.

**Table 3.** Mixed state reconstruction.

| Contribution of components | Entropy, $S$ | | Fidelity, $F\left(\rho_{theor}^{mixed}, \rho_{exp}^{comp}\right)$ | |
|---|---|---|---|---|
| | Single plate (63% mixture) | Double plate (98% mixture) | Single plate (63% mixture) | Double plate (98% mixture) |
| 1-7 | 0.6344 | 0.9835 | 0.9974 | 0.9990 |
| 2-6 | 0.5310 | 0.9562 | 0.9934 | 0.9975 |
| 3, 4, 5 | 0.2906 | 0.7150 | 0.9617 | 0.9402 |
| 2, 4, 6 | 0.6202 | 0.9871 | 0.9973 | 0.9993 |
| 1-4 | 0.3820 | 0.7870 | 0.9537 | 0.9479 |
| 2, 3, 7 | 0.5214 | 0.4521 | 0.8785 | 0.7860 |

**8. Influence of instrumental uncertainties due to mechanically induced optical anisotropy on the polarization reconstruction**

One of the possible uncertainties arising at the stages of preparation transformation and measurement of polarization states is caused by piezoelectric effect (or "photoelasticity") [26].

As an example we show the photograph pictures (figure 8) of initially isotropic glass plate (1" diameter substrate for the regular dielectric mirror) settled between crossed polarization prisms (figure 9).

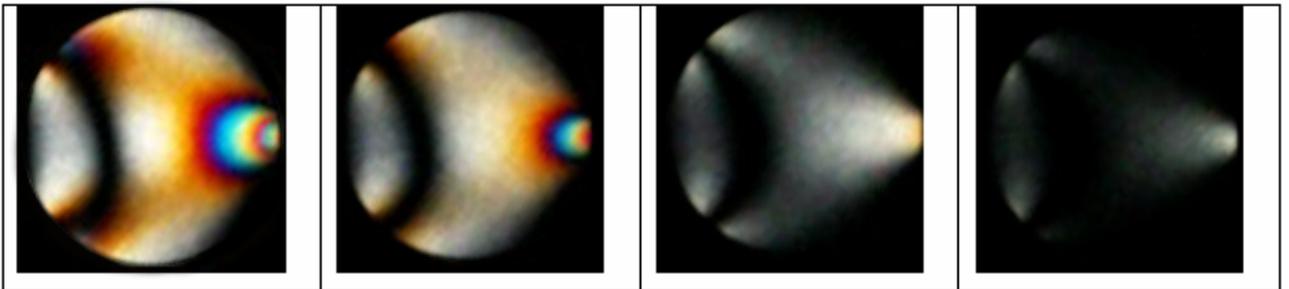

Figure 8. Photograph pictures of initially isotropic glass plate (1" diameter substrate for the regular dielectric mirror) settled between crossed polarization prisms (figure 9). The pictures correspond to different strength (increase from the right to the left) of the mechanical force $F$ applied to the plate $Q$. The force is applied at approximately $90^0$ with respect to vertical direction.

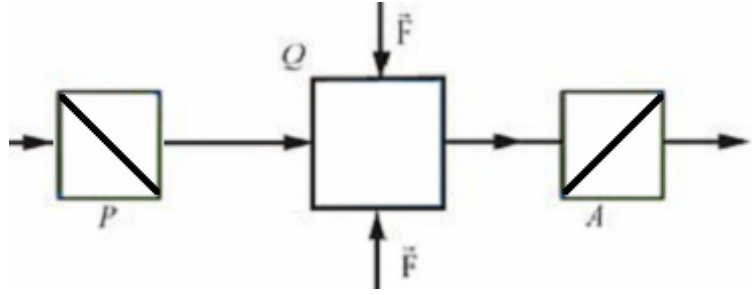

Figure 9. Set-up for observing polarization transformations caused by induced anisotropy. The fused-silica plate *Q* to which a mechanical stress *F* is applied is placed between crossed polarizing prisms *P* and *A*.

The 1-inch diameter glass plate is settled in regular holder (Thorlabs, KM100). The figure demonstrates clearly that the appearance of artificial anisotropy caused unwanted polarization transformation in the plate. Analogous effect can be easily observed in the reflected light if instead of the plate the regular dielectric mirror is used. If the effect is linear with respect to applied mechanical stress $\sigma = \frac{F}{S}$, then the following relation holds:

$$\Delta n = n_e - n_o = K_1 \sigma, \tag{55}$$

where $K_1$ is the Brewster constant with typical value about $K_1 \approx 10^{-12} \div 10^{-11}\ m^2/Newton$.

If the stress is applied uniformly then the optical elements takes birefringence and behaves as an uniaxial crystal with optical axis parallel to the mechanical force.

To increase quality of the state reconstruction and preparation these kinds of uncertainties have to be taken into account. However to do this one needs to know which particular transformation is performed by given element. In this particular case direct calculation of opto-mechanical effect meets some trouble and seems to be ineffective. That is why we have used quantum process tomography (QPT) method to reconstruct the transformation matrix.

As a set of input states we used "R4" set (see above), formed tetrahedron on the Poincaré - Bloch sphere. Each state subjected to polarization transformation induced by mechanical stress was measured according to R4 protocol. As a result the $\chi$-matrix (which completely describes the quantum process) was reconstructed.

The experimental set-up is completely analogous to those one shown on the figure 6a, however instead of thick quartz plates we use 1-inch glass plate placed in the standard Thorlabs holder. Using small pinholes to limit the beam we select particular region on the plate where homogeneous distribution of mechanical stress (or colour on figure 8) was present. In our experiment we selected the central region of the plate. This simple technique guaranties the spatial homogeneity of the selected region which allows fixed transformation matrix to be measured" или "which leads to fixed transformation matrix being measured. We prepare the set of initial states using two plates with $824\mu m$ and $356\mu m$ thickness and initial vertically polarized light. Then we performed analogous measurement with protocol B36 using the plate $312.7\mu m$. The reconstructed Choi- Jamiolkowski state has the following form:

$$\rho_\chi = \begin{pmatrix} 0.50211 & 0.01813 + 0.00096273i & -0.009349 - 0.015417i & 0.19478 + 0.45978i \\ 0.01813 - 0.00096273i & 0.00065645 & -0.00036712 - 0.00053873i & 0.0079146 + 0.016228i \\ -0.009349 + 0.015417i & -0.00036712 + 0.00053873i & 0.00064743 & -0.017744 - 0.0025803i \\ 0.19478 - 0.45978i & 0.0079146 - 0.016228i & -0.017744 + 0.0025803i & 0.49659 \end{pmatrix}$$

Figure 10 illustrates the results of statistical reconstruction of the process. We see good agreement between the statistical model and experimental data.

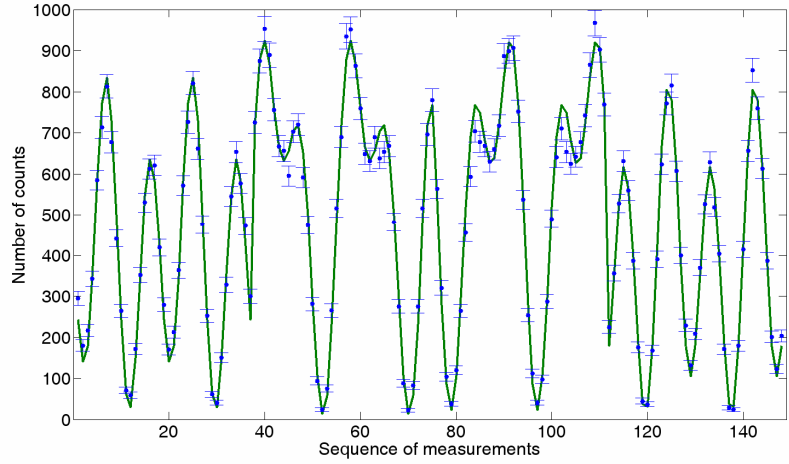

Figure 10. Statistical reconstruction of the polarization transformation caused by photoelasticity. Dots - the observed number of counts, line - the expected number of counts computed using the reconstructed $\chi$-matrix.

To check the homogeneity of induced anisotropy we reconstruct the density matrix for R4 set and calculated entropy

$$S = -\sum_{n=1}^{2} \lambda_n \log_2 \lambda_n$$

with $\lambda$ being the eigen values of the density matrix. The degree of mixture did not exceed 0.1% which confirms an adequacy of our method.

The reconstructed $\chi$-matrix clearly shows that the glass plate subjected to induced anisotropy transforms each of the state from R4 set uniformly, i.e. can be presented as standard SU2 matrix:

$G = \begin{bmatrix} t & r \\ -r^* & t^* \end{bmatrix}$, where $t = \cos\delta + i\sin\delta\cos(2\alpha)$, $r = i\sin\delta\sin(2\alpha)$, and $\delta = \frac{\pi\Delta nL}{\lambda}$ is an optical thickness.

Knowing both crucial parameters: the wavelength $\lambda$ and geometrical thickness of the plate L one can reconstruct the orientation of induced optical axis $\alpha$ and birefringence $\Delta n$. In our case for particular mechanical stress applied to the glass plate we got the following meanings. $\alpha = 91°, \Delta n = 2.2 \times 10^{-3}$.

We would like to note that considered distortion of polarization states becomes essential if relatively small mechanical stress is applied to the initially isotropic optical elements (like beamsplitters, substrates of dielectric mirrors, glass filters, etc.). Usually it happens when mentioned above optical elements are settled in the holders.

Practically avoiding this sort of effects meets some trouble since even tiny mechanical pressure causes unwanted anisotropy which introduces unpredictable changes in polarization state passing through such an element. Moreover these effects become crucial at the stage of quantum polarization states are prepared or measured. That is why developed QPT methods for testing different parts of corresponding apparatus seem to be effective and practically required in quantum computation/communication systems.

## 9. Conclusions

In this work we present common approach to the statistical reconstruction of the quantum process. The approach is based on $\chi$-matrix and Choi-Jamiolkowski states and includes well-developed methods of quantum state tomography applied to the Choi-Jamiolkowski state(s). Performed experiments clearly demonstrate adequacy of the suggested QPT method to analysis of polarization transformations performed in anisotropic and dispersive media. We hope that the relevant formalism will prove to be extremely important to analyze the quality of the designed quantum gates.

## 10. Acknowledgments
We would like to thank A.N. Korotkov and L.V. Belinsky for many helpful discussions. This work was supported in part by Russian Foundation of Basic Research (projects 10-02-00204a, 10-02-00414-a), by Program of Russian Academy of Sciences in fundamental research, and by Council on grants of the President of the Russian Federation (project MK-4277.2011.2). E V Moreva is grateful to the Dynasty Foundation for financial support.